\documentclass[12pt]{svmult}

\newcommand{\be}{\begin{equation}}
\newcommand{\ee}{\end{equation}}
\newcommand{\bea}{\begin{eqnarray}}
\newcommand{\eea}{\end{eqnarray}}

\usepackage{xcolor}
\usepackage[colorlinks=true, pdfstartview=FitV,linkcolor=blue, citecolor=blue, urlcolor=blue]{hyperref}
\usepackage{psfrag}
\usepackage{latexsym,array,theorem,mathrsfs,subfigure,bm,float}
\usepackage{amsfonts}
\usepackage{amssymb}

\begin{document}

\title*{Hairy black holes in theories with massive gravitons}

\author{Mikhail S. Volkov}

\institute{M.S.Volkov
\at Laboratoire de Math\'{e}matiques et Physique Th\'{e}orique CNRS-UMR 7350, \\
Universit\'{e} de Tours, Parc de Grandmont, 37200 Tours, FRANCE;
\vspace{0.5 mm}
\at Institut des Hautes Etudes Scientifiques (IHES), 91440 Bures-sur-Yvette, FRANCE;
\vspace{0.5 mm}
\at Department of General Relativity and Gravitation, Institute of Physics,\\
Kazan Federal University, Kremlevskaya str.18, 420008 Kazan, RUSSIA.
\vspace{0.5 mm}
\at \email{volkov@lmpt.univ-tours.fr}
}

\maketitle

\abstract{
This is a brief survey of the known black hole solutions 
in the theories of ghost-free bigravity and massive gravity. 
Various black holes exist in these theories, in particular 
those supporting a massive graviton hair. 
However, it seems that solutions 
which could be astrophysically relevant are the same as in General
Relativity, or very close to them. 
Therefore, the no-hair conjecture essentially applies, and so  
it would be hard to detect 
the graviton mass by observing black holes. 
}

\section{Black holes and the no-hair conjecture}

More than 40 year ago J.A.~Wheeler summarized the progress 
in the area of black hole physics at the time by his famous phrase: 
{\it  black holes have no hair} \cite{Ruffini}. More precisely, this means that   

\begin{itemize} 
\item 
\textcolor{black}{
All stationary black holes are completely characterized by their mass,
angular momentum, and electric charge measurable from infinity.}

\item
Black holes cannot support \textcolor{black}{hair} =
external fields distributed 
close to the horizon but not seen from infinity.

\end{itemize}
Therefore, 
according to the ho-hair conjecture, 
the only allowed characteristics of stationary black holes are 
those associated with the Gauss law. 
The logic behind this is the following.
Black holes are formed in the gravitational collapse, which 
is so violent a process  that it breaks all the usual   
conservation laws not related to the exact symmetries.   
For example, the chemical content, the baryon number, etc. are not conserved
during the collapse -- the black hole `swallows' all the memory of them.    
Everything that can be absorbed by the black hole gets  absorbed. 
Only few exact local symmetries, 
such as the local Lorentz or local U(1), can survive
the gravitational collapse. Associated to them conserved quantities -- the mass,
angular momentum, and electric charge -- cannot be absorbed by the  
black hole and remain attached to it as parameters. They give rise 
to the Gaussian  fluxes that can be measured at infinity. 

The no-hair conjecture essentially implies  that 
the only asymptotically flat
black holes in Nature should be 
those described by the Kerr-Newman solutions.  
And indeed, a number of {\sl the uniqueness theorems  }
\cite{Israel:1967wq,Carter,Mazur:1982db} confirm that all  
stationary and asymptotically flat {\sl electrovacuum} black holes 
with a non-degenerate horizon should 
belong to the Kerr-Newman family. 

The electrovacuum uniqueness theorems do not directly apply to systems 
with matter fields other than the electromagnetic field. 
The field equations for such systems read schematically 
\be                                      \label{1a}
G_{\mu\nu}=8\pi G T_{\mu\nu}(\Psi),~~~~~\Box\Psi=V(\Psi),
\ee
where $\Psi$ denotes the matter field, or several interacting matter fields. 
 One can wonder if these equations 
admit asymptotically flat black hole solutions with the curvature bounded everywhere 
outside the black hole horizon.  According to the no-hair conjecture,
the answer should be negative, but to prove this requires considering 
each matter type separately.  In view of this, a number of {\sl the no-hair theorems}
have been proven to confirm  the absence of static black hole solutions 
of Eqs.(\ref{1a}) in the cases where  $\Psi$ denotes  
scalar, spinor, etc. fields 
\cite{
Bekenstein:1972ny,
Bekenstein:1971hc,
Bekenstein:1972ky,
Bekenstein:1995un,
Mayo:1996mv,
Bekenstein:1996pn}. 
The common feature in all these cases is that if $\Psi$ does not vanish, then 
the field equations require that it should 
diverge at the black hole horizon, where the curvature diverges too.  
Therefore, to get regular black holes one is bound to set $\Psi=0$, 
but then the solution is a vacuum black hole belonging to the Kerr-Newman family%
\footnote{It has recently been shown that these arguments can be circumvented 
for fine-tuned black hole mass and angular momentum \cite{Hod:2012px}. 
This allows one to construct spinning hairy black holes  
which do not admit a static limit \cite{Herdeiro:2014goa}.
}. 
All this confirms the non-existence  of {\sl hairy} black holes.

The first explicit evidence against the no-hair conjecture was found 
20 years after its formulation,   in the 
context of the Einstein-Yang-Mills theory with gauge group SU(2). 
This theory contains all the electrovacuum solutions, hence  all 
Kerr-Newman black holes \cite{Yasskin:1975ag}, 
because the electromagnetic U(1) gauge group 
is contained in SU(2). However, it also admits
static black holes supporting a non-trivial Yang-Mills field 
which asymptotically decays as $1/r^3$, so that 
the corresponding Gaussian flux is zero \cite{Volkov:1989fi,Volkov:1990sva}. 
Close to the horizon the geometry deviates from the 
Schwarzschild one, but the deviations rapidly decay with distance
and cannot be seen from infinity. Therefore, such black holes support a {\sl hair}.  

Subsequent developments have revealed that the Einstein-Yang-Mills black holes 
can be generalized to include  scalar fields, as for example a 
Higgs field, which leads to a variety of new
solutions describing hairy black holes  \cite{Volkov:1998cc}. 
In particular, it turns out that regular gravitating solitons, as for example gravitating 
magnetic monopoles or gravitating Skyrmions, can be generalized to contain inside 
a small black hole. This gives rise to black holes with a `solitonic hair'. 
However, when the black hole size 
exceeds a certain critical value, the black hole `swallows the soliton'
and `looses its hair', becoming a
Kerr-Newman black hole \cite{Volkov:1998cc}. 

Yet more hairy black holes can be obtained in models inspired by string theory and 
including a dilaton \cite{Kleihaus:1997ws}, the curvature corrections and so on 
\cite{Volkov:1998cc}. 
Adding a cosmological term, positive or negative, gives asymptotically 
(anti)-de Sitter hairy black holes \cite{Gubser:2008wv}. 
Summarizing, one can say that hairy black holes arise generically  in 
physical models. However, large hairy black holes are typically unstable 
and loose the hair when perturbed, 
whereas the stable ones  are typically  very small \cite{Volkov:1998cc}. 
As a result, despite a large number of solutions describing hairy 
black holes in various systems, 
it seems that the no-hair conjecture essentially holds
for the  astrophysical black holes, all of which 
should be of the Kerr-Newman type.

In what follows we shall be considering black holes 
in theories with massive gravitons -- the ghost-free
bigravity and massive gravity. Some of 
these black holes are of 
the known Kerr-Newman(-de Sitter) type, but there are also 
black holes supporting a massive graviton hair. 
However, the hairy black holes turn out to be either 
asymptotically anti-de Sitter (AdS),
or cosmologically large, which contradicts the observations. 
Therefore, the
astrophysical black holes should be described by the Kerr-Newman(-de Sitter) metrics,
possibly with small corrections in the near-horizon region, 
so that the no-hair conjecture
essentially holds.

\section{Theories with massive gravitons}

The idea that gravitons could have a tiny mass was proposed long ago \cite{Fierz:1939ix},
but it attracted a particular interest  after the recent 
discovery of the special massive gravity theory 
by de Rham, Gabadadze, and Tolley (dRGT) 
\cite{deRham:2010kj} 
(see \cite{Hinterbichler:2011tt},\cite{deRham:2014zqa}
for a review).
Before this discovery it had been known that the massive gravity theory
generically  
had six  propagating degrees of freedom (Dof). Five of them could be 
associated with the polarizations of the massive graviton, while the sixth 
one,  usually called Boulware-Deser (BD) ghost, 
is unphysical, because 
it has a negative kinetic energy and  renders the whole theory unstable 
\cite{Boulware:1973my}. 
The specialty of the dRGT theory is that it  
contains two Hamiltonian constraints 
 which eliminate one of the six Dof 
\cite{Hassan:2011hr},%
\cite{Hassan:2011ea},%
\cite{Kluson:2012wf},%
\cite{Comelli:2012vz},%
\cite{Comelli:2013txa}.
Therefore, there remain just the right 
number of Dof to describe massive gravitons and
so the theory is referred to as ghost-free.  
This does not mean that all solutions are 
stable in this theory, since there could be other instabilities,
which should be checked in each particular case. However, since the 
most dangerous BD ghost instability is absent, the theory of 
\cite{deRham:2010kj}
and its bigravity generalization \cite{Hassan:2011zd}
can be considered as healthy  physical models 
for interpreting the observational data.

These theories can be used to  
explain the current cosmic acceleration 
\cite{1538-3881-116-3-1009,0004-637X-517-2-565}. 
This acceleration could be accounted for by introducing a cosmological term in
Einstein equations, however, this would pose the problem of explaining the origin
and value of this term. An alternative possibility is to consider 
{modifications}
of General Relativity (GR), and theories with massive gravitons are {natural}
candidates for this, since the graviton mass can effectively manifest itself
 as a small cosmological term \cite{PhysRevD.66.104025}.

Theories with massive gravitons are described by two metrics, $g_{\mu\nu}$
and $f_{\mu\nu}$. In massive gravity theories the f-metric is non-dynamical
and is usually chosen to be flat, although other choices are also possible,
while the dynamical g-metric describes massive gravitons.   
In bigravity theories \cite{Hassan:2011zd} both metrics are dynamical and describe together 
two gravitons, one massive and one  massless.
The theory contains two gravitational couplings, $\kappa_g$ 
and $\kappa_f$,
and in the $\kappa_f\to 0$ limit the f-metric decouples 
and can be chosen to be flat. Therefore, the bigravity theory is more general,
while the massive gravity theory can be viewed as its special case.

All known bigravity black holes were obtained in Ref.\cite{Volkov:2012wp}
(see also \cite{Comelli:2011wq}),
with the exception of special solutions discovered in Ref.\cite{Brito:2013xaa}. 
These black holes 
can be divided into three types. First,
there are solutions for which the two metrics are proportional, 
$f_{\mu\nu}=C^2g_{\mu\nu}$ with a constant $C$,
where $g_{\mu\nu}$ fulfills the Einstein equations 
with  a cosmological term $\Lambda(C)\propto m^2$. 
If $C=1$ then  $\Lambda=0$
and one obtains all solutions of the vacuum GR, in particular 
the vacuum black holes. For other values of $C$ one has $\Lambda(C)\neq 0$,
which gives rise to black holes with a cosmological term.  
None of these solutions
fulfill equations of the massive gravity 
theory with a flat f.

Secondly, imposing spherical symmetry, there are black holes 
described by two metrics which are not simultaneously diagonal. 
They formally decouple one from the other and 
each of them fulfills its own set of Einstein equations with its own
cosmological term.  The g-metric is Schwarzschild-de Sitter, whereas
the f-metric can be chosen to be AdS,
with $\Lambda_f\sim \kappa_f^2$, and it becomes flat when $\kappa_f\to 0$,
in which limit the dRGT massive gravity is naturally recovered. Therefore, 
these solutions  
exist {both} in the bigravity and dRGT massive gravity theories. 
In the latter case they exhaust  all known 
black hole solutions.

Solutions of the  third type are obtained when 
the two metrics are both diagonal but not proportional. 
One obtains in this case more complex solutions describing 
static black holes with a massive graviton hair, which can be either 
asymptotically AdS \cite{Volkov:2012wp}, or  asymptotically flat \cite{Brito:2013xaa},
although in the latter case their size should be comparable with the Hubble radius.

A more detailed description of the currently known 
bigravity and massive gravity black holes is given below.

\section{Ghost-free bigravity}
The theory of the ghost-free bigravity \cite {Hassan:2011zd} 
is defined on a four-dimensional spacetime manifold
equipped with two metrics, ${g_{\mu\nu}}$ and 
${f_{\mu\nu}}$, which describe two interacting gravitons, one massive 
and one massless. 
The kinetic term for each metric is chosen to be of the 
standard Einstein-Hilbert form, 
while the interaction between them is 
described by a local potential ${\cal U}[g,f]$ which does not contain 
derivatives and is expressed by  
 a scalar function 
of the tensor 
\be                             \label{gam}
\gamma^\mu_{~\nu}=\sqrt{{g}^{\mu\alpha}{f}_{\alpha\nu}}.
\ee
Here ${g}^{\mu\nu}$ is the inverse of ${g}_{\mu\nu}$
and the square root
is understood in the matrix sense, i.e. 
\be                     \label{gamgam}
(\gamma^2)^\mu_{~\nu}\equiv \gamma^\mu_{~\alpha}\gamma^\alpha_{~\nu}={g}^{\mu\alpha}
{f}_{\alpha\nu}.
\ee 
The action is (with the metric signature $-+++$) 
\bea                                      \label{1}
S[{g},{f}]
&=&\frac{1}{2\kappa_g^2}\,\int d^4x\, \sqrt{-{g}}\,R({g})
+\frac{1}{2\kappa_f^2}\,\int d^4x\, \sqrt{-{f}}\, {\cal R}({f})  \nonumber \\
&-&\frac{m^2}{\kappa^2}\int d^4x\, \sqrt{-{g}} \, {\cal U}[{g},{f}] \,
\,, 
\eea
where $R$ and ${\cal R}$ are the Ricci scalars for ${g}_{\mu\nu}$ and 
${f}_{\mu\nu}$, respectively,  $\kappa_g^2=8\pi G$ and $\kappa_f^2=8\pi {\cal G}$
 are the corresponding gravitational couplings, while 
$\kappa^2=\kappa_g^2+\kappa_f^2$ and 
$m$ is the graviton mass.
The interaction 
between the two metrics is given by
\be                             \label{2}
{\cal U}=\sum_{k=0}^4 b_k\,{\cal U}_k(\gamma),
\ee
where $b_k$ are parameters, while  
${\cal U}_k(\gamma)$ are defined 
by the 
relations
\bea                        \label{UU}
{\cal U}_0(\gamma)&=&1,~~~~
{\cal U}_1(\gamma)=
\sum_{A}\lambda_A=[\gamma],\nonumber \\
{\cal U}_2(\gamma)&=&
\sum_{A<B}\lambda_A\lambda_B 
=\frac{1}{2!}([\gamma]^2-[\gamma^2]),\nonumber \\
{\cal U}_3(\gamma)&=&
\sum_{A<B<C}\lambda_A\lambda_B\lambda_C
=
\frac{1}{3!}([\gamma]^3-3[\gamma][\gamma^2]+2[\gamma^3]),\nonumber \\
{\cal U}_4(\gamma)&=&
\lambda_0\lambda_1\lambda_2\lambda_3
=
\frac{1}{4!}([\gamma]^4-6[\gamma]^2[\gamma^2]+8[\gamma][\gamma^3]+3[\gamma^2]^2
-6[\gamma^4])\,. 
\eea
Here $\lambda_A$ ($A=0,1,2,3$) are the eigenvalues of $\gamma^\mu_{~\nu}$, 
and, using the hat to denote matrices, one has defined 
$[\gamma]={\rm tr}(\hat{\gamma})\equiv \gamma^\mu_{~\mu}$, 
$[\gamma^k]={\rm tr}(\hat{\gamma}^k)\equiv (\gamma^k)^\mu_{~\mu}$. 
The (real) parameters $b_k$ could be arbitrary, however, if one requires 
flat space to be a solution of the theory, and $m$ to be the Fierz-Pauli 
mass of the graviton \cite{Fierz:1939ix}, 
then the five $b_k$'s are expressed in terms of 
{\it two} free parameters $c_3,c_4$ as follows: 
\bea                \label{bbb}
b_0&=&4c_3+c_4-6,~~b_1=3-3c_3-c_4,~~
b_2=2c_3+c_4-1,~~\nonumber \\
b_3&=&-(c_3+c_4),~~
b_4=c_4.
\eea
The theory (\ref{1}) propagates 7=5+2 Dof corresponding to 
the polarizations of two gravitons, one massive and one massless. 
Before this theory was discovered 
\cite{Hassan:2011zd}, more general bigravity models, sometimes called f-g theories,
had been considered \cite{Isham:1971gm}. In these models  
the potential ${\cal U}$ is a scalar function of 
$H^\mu_{~\nu}=\delta^\mu_\nu
-g^{\mu\alpha}f_{\alpha\nu}$ of the form 
\be                            \label{PF}
{\cal U}= \frac18\,
(H^{\mu}_{~\nu}H^{\nu}_{~\mu}-(H^\mu_{~\mu})^2)
+\ldots ,
\ee 
where 
the dots denote all possible 
higher order scalars made of $H^\mu_{~\nu}$. A particular choice of these terms
leads to (\ref{2}). The generic f-g theories 
propagate 7+1 Dof, 
the additional one being the BD ghost \cite{Boulware:1973my}.

Introducing the mixing angle $\eta$ such that
$\kappa_g=\kappa\cos\eta$, $\kappa_f=\kappa\sin\eta$ and 
varying the action (\ref{1})  gives the field equations 
\bea                                  \label{Einstein}
G^\mu_\nu&=&m^2\cos^2\eta\, T^{\mu}_{~\nu}\,,~~~~~~~ \\
{\cal G}^\mu_\nu&=&m^2\sin^2\eta\, {\cal T}^{\mu}_{~\nu}\,, \label{Einstein1}
\eea
where $G^\mu_\nu$ and ${\cal G}^\mu_\nu$ are the  Einstein tensors for $g_{\mu\nu}$
and $f_{\mu\nu}$. 
The graviton energy-momentum tensors obtained by varying the interaction 
${\cal U}$ are 
\bea                        \label{T}
&&
T^{\mu}_{~\nu}=
\,\tau^\mu_{~\nu}-{\cal U}\,\delta^\mu_\nu,~~~~~
{\cal T}^{\mu}_{~\nu}
=-\frac{\sqrt{-g}}{\sqrt{-f}}\,\tau^\mu_{~\nu}\,,
\eea
where 
\bea                                \label{tau}
\tau^\mu_{~\nu}&=&
\{b_1\,{\cal U}_0+b_2\,{\cal U}_1+b_3\,{\cal U}_2
+b_4\,{\cal U}_3\}\gamma^\mu_{~\nu} \nonumber \\
&-&\{b_2\,{\cal U}_0+b_3\,{\cal U}_1+b_4\,{\cal U}_2\}(\gamma^2)^\mu_{~\nu} \nonumber  \\
&+&\{b_3\,{\cal U}_0+b_4\,{\cal U}_1\}(\gamma^3)^\mu_{~\nu} \nonumber \\
&-&b_4\,{\cal U}_0\,(\gamma^4)^\mu_{~\nu}\,,
\eea
with ${\cal U}_k\equiv {\cal U}_k(\gamma)$. 
The Bianchi identities for (\ref{Einstein}) and (\ref{Einstein1})  imply that 
\be                                   \label{Bian} 
\stackrel{(g)}{\nabla}_\mu\! T^{\mu}_{~\nu}=0,
~~~~~
\stackrel{(f)}{\nabla}_\mu\!{\cal T}^{\mu}_{~\nu}
=0,
\ee 
where $\stackrel{(g)}{\nabla}$ and $\stackrel{(f)}{\nabla}$
are the covariant derivatives with respect to $g_{\mu\nu}$ and 
$f_{\mu\nu}$. In fact, the latter of these
conditions is not independent and follows from 
the former one 
 in view of the diffeomorphism invariance of 
the interaction term.  

If $\eta\to 0$ and $\sin^2\eta\,{\cal T}^\mu_{~\nu}\to 0$, then 
equations (\ref{Einstein1}) for the f-metric decouple and 
their solution enters the g-equations (\ref{Einstein}) as a fixed reference metric.
The g-equations describe in this case a massive gravity theory.  
If f becomes
flat for $\eta\to 0$, then one recovers the dRGT theory \cite{deRham:2010kj}. 
Therefore, the massive gravity theory is contained in the bigravity.

\section{Proportional backgrounds \label{GR}}

The simplest solutions of the bigravity equations are obtained
by assuming the two
metrics to be  proportional \cite{Volkov:2012wp},\cite{Volkov:2013roa}, 
\be
f_{\mu\nu}=C^2 g_{\mu\nu}. 
\ee
The energy-momentum tensors (\ref{T}) then become
\be
T^{\mu}_{~\nu}=-\Lambda_g(C) \delta^\mu_{~\nu},~~~
{\cal T}^{\mu}_{~\nu}=-\Lambda_f(C) \delta^\mu_{~\nu}\,,
\ee
with 
\begin{eqnarray}                  \label{LAM}
\Lambda_g(C) &=&m^2\cos^2\eta
\left(b_0 +3 b_1\,C+3 b_2\,C^2
+ b_3\,C^3\right)\,,~
\label{Lmbd}  \nonumber 
\\
\Lambda_f(C) &=&m^2\,{\sin^2\eta\over C^3}
\left(b_1 +3 b_2 C\,+3 b_3 C^2\,
+ b_4 C^3\,\right)\,.
\end{eqnarray}
Since  the energy-momentum tensors should be conserved, 
it follows that $C$ is a constant.
As a result, 
one obtains two sets of Einstein equations,
\bea                           \label{g+f}
G_\mu^\nu   +\Lambda_g(C) \delta_\mu^\nu=0\,,~~~~
{\cal G}_\mu^\nu    +\Lambda_f(C) \delta_\mu^\nu=0
\,.
\eea
Since one has 
${\cal G}_\mu^\nu=  G_\mu^\nu/C^2 $, 
it follows that $\Lambda_f =  \Lambda_g/C^2$,
which gives an algebraic equation for $C$. 
If the parameters  $b_k$ are chosen  
according to 
Eq.(\ref{bbb}), then this equation reads
\bea  \label{LAMC}
0=(C-1)[(c_3+c_4)C^3+(3-5c_3+(\chi-2)c_4)C^2~~~~\nonumber   \\
+((4-3\chi)c_3+(1-2\chi)c_4-6)C+(3c_3+c_4-1)\chi],
\eea
with $\chi=\tan^2\eta$, while the cosmological constant is
\be
\frac{\Lambda_g}{m^2\cos^2\eta}=(1-C) 
((c_3+c_4)C^2+(3-5c_3-2c_4)C+4c_3+c_4-6). 
\ee
Depending on values of $c_3,c_4,\eta$, 
Eq.(\ref{LAMC}) can have up to four real roots,
so that there can be solutions with four different
values of the cosmological constant, which can be positive, negative, or zero. 

One solution of (\ref{LAMC}) is
$C=1$, in which case the 
two metrics coincide, 
$g_{\mu\nu}=f_{\mu\nu}$, while
$\Lambda_g=0$, so that 
the vacuum GR is recovered. 
Therefore, the black hole solutions obtained in this case 
are either Kerr, or Kerr-de Sitter, or Kerr-AdS. 
None of these solutions admit
the massive gravity limit with a flat f-metric.

\section{Solutions with non-bidiagonal metrics \label{off}}

Let us assume both metrics
to be invariant under spatial SO(3) rotations.
Since the theory is invariant under diffeomorphisms, one can 
choose the spacetime coordinates such that the g-metric is diagonal. 
However, the f-metric will  in general contain an off-diagonal term,
so that the two metrics can be parameterized as 
\bea                             \label{ansatz}
ds_g^2&=&-N^2dt^2+\frac{dr^2}{\Delta^2}+R^2d\Omega^2\,, \nonumber \\
ds_f^2&=&-\left(aNdt+\frac{c}{\Delta}\,dr\right)^2
+\left(cNdt-\frac{b}{\Delta}\,dr\right)^2+u^2R^2d\Omega^2\,,
\eea 
with $d\Omega^2=d\vartheta^2+\sin^2\vartheta d\varphi^2$\,. 
The amplitudes 
$N,\Delta,R$ depend on $r$, while  $a,b,c,u$ can in general depend  on $t,r$.
It is straightforward to check that the matrix square root is 
\be                                  \label{gamma}
\gamma^\mu_{~\nu}=\sqrt{g^{\mu\alpha}f_{\alpha\nu}}=\left(
\begin{array}{cccc}
a & c/(\Delta N) & 0 & 0 \\
-c\Delta N & b & 0 & 0 \\
0 & 0 & u & 0 \\
0 & 0 & 0 & u
\end{array}
\right)\,,
\ee
whose eigenvalues are 
\be                              \label{lambda}
\lambda_{0,1}=\left.\left.\frac12\right(a+b\pm\sqrt{(a-b)^2-4c^2}\right),
~~~\lambda_2=\lambda_3=u.
\ee
Inserting this  to (\ref{UU}) gives 
\bea
{\cal U}_1&=&a+b+2u,~~~~~{\cal U}_2=u(u+2a+2b)+ab+c^2\,, \nonumber \\
{\cal U}_3&=&u\,(au+bu+2ab+2c^2),~~~~~
{\cal U}_4=u^2(ab+c^2). 
\eea
Although the eigenvalues (\ref{lambda}) can be complex-valued, 
the ${\cal U}_k$'s are always real. 
It is straightforward to compute the energy-momentum tensors $T^\mu_{~\nu}$
and ${\cal T}^\mu_{~\nu}$ defined by Eqs.(\ref{T}),(\ref{tau}). In particular,
one finds 
\be                              \label{A20}
T^0_{~r}=\frac{c}{\Delta N}\,[b_1+2b_2u+b_3u^2].
\ee
Since the g-metric is static,
there is no radial energy flux, and so 
$T^0_{~r}$ should be zero. Therefore,    
either $c$ should vanish, or the expression in brackets in (\ref{A20}) vanishes.  
The former option will be considered in the next Section, 
while presently let us assume that $c\neq 0$ and
\be                         \label{bb}
b_1+2b_2u+b_3u^2=0.
\ee
This yields
\be                                        \label{u}
u=\frac{1}{b_3}\left( 
-b_2\pm\sqrt{b_2^2-b_1 b_3}
\right).
\ee
Notice that $u$ was a priori a function of $t,r$, but now it is restricted to be a constant.  
Using this, one finds that  $T^0_{~0}=T^r_{~r}=-\lambda_g$ and 
${\cal T}^0_{~0}={\cal T}^r_{~r}=-\lambda_f$ where 
\be
\lambda_g=b_0+2b_1u+b_2u^2,~~~~~~\lambda_f=
\frac{b_2+2b_3u+b_4u^2}{u^2}.
\ee
The conditions 
$
\stackrel{(g)}{\nabla}_\rho T^{\rho}_{\lambda}=0\,
$
reduce in this case to the requirement that 
$T^0_{~0}-T^\vartheta_{~\vartheta}$ should vanish. 
On the other hand, one finds 
\be
T^0_{~0}-T^\vartheta_{~\vartheta}=(b_2+b_3 u)
[(u-a)(u-b)+c^2],  \label{cons}
\ee
and since this has to vanish,
either the first or the second factor on the right should be zero. 
Let us assume that one of these conditions is fulfilled. 
Then one has
$T^0_{~0}=T^\vartheta_{~\vartheta}$ and ${\cal T}^0_{~0}
={\cal T}^\vartheta_{~\vartheta}$,
hence both energy-momentum tensors are 
proportional to the unit tensor,
$T^\mu_{~\nu}=-\lambda_g\delta^\mu_\nu$ and 
${\cal T}^\mu_{~\nu}=-{\lambda_f}\delta^\mu_\nu$. 
The field equations (\ref{Einstein})  then 
reduce to 
\bea
G^\rho_\lambda+\Lambda_g \delta^\rho_\lambda
&=&0 \,,\label{ee1} \\
{\cal G}^\rho_\lambda+
{\Lambda}_f \delta^\rho_\lambda&=&0 \,,       \label{ee2}
\eea
where 
\be                                 \label{lam}
\Lambda_g=m^2\cos^2\eta\,\lambda_g,~~~~~
 {\Lambda}_f=m^2\sin^2\eta\,{\lambda}_f. 
\ee
As a result, 
the two metrics decouple one from the other,
and the graviton mass gives rise to the two cosmological terms. 
 If the parameters $b_k$ are chosen
according to (\ref{bbb}), then $\lambda_g+u^2\lambda_f=-(u-1)^2$, 
therefore, if $\Lambda_g>0$ then 
$\Lambda_f<0$. 

Since we want the g-metric to describe a black hole geometry, 
the solution of (\ref{ee1}) is the Schwarzschild-de Sitter
metric. 
On the other hand,  as
the cosmological term for the f-metric is negative, the solution 
of (\ref{ee2}) can be chosen to be 
AdS. Therefore,
\bea                                \label{gf}
ds_g^2&=&-\Sigma(r)\, dt^2+\frac{dr^2}{\Sigma(r)}
+r^2d\Omega^2\,,~~~~~~\Sigma(r)=1-\frac{2M}{r}-\frac{{\Lambda_g}}{3}\,r^2
,\nonumber \\
ds_f^2&=&-{\cal D}(U)\, dT^2+\frac{dU^2}{{\cal D}(U)}
+U^2d\Omega^2\,,~~~~~~{\cal D}(U)=1-\frac{{\Lambda_f}}{3}\,U^2, 
\eea
with $U=ur$. 
It is worth noting that, since 
$\Lambda_f\sim\sin^2\eta\to 0$ 
when $\eta\to 0$, the f-metric becomes flat in this limit. 
Therefore, the 
solutions apply both in the bigravity theory and in the dRGT massive gravity. 

\subsection{Imposing the consistency condition}

The solution (\ref{gf})
is not yet complete, since the two metrics  are expressed in two different
coordinate systems, $t,r$ and $T,U$, whose relation to each other is not known. 
One has $U=ur$ 
but the function $T(t,r)$ is still undetermined.  
We therefore remember that up to now we have not considered the consistency 
condition, which requires that the expression in (\ref{cons}) should vanish. 
This condition will be fulfilled in either  of the following two cases:
\bea                     \label{cons1}
\mbox{I:}&&~~~~~(b_2+b_3 u)=0;\\                
\mbox{II:}&&~~~~~(u-a)(u-b)+c^2=0.  \label{cons2}
\eea
In case {I}, since $u$ is already 
expressed in terms of $b_1,b_2,b_3$ by Eq.(\ref{u}), the condition (\ref{cons1}) 
imposes a constraint on values of these parameters. Therefore, this condition 
is possible only for the special subclass of the theory 
characterized by the restricted values of $b_k$. 
Within this subclass the consistency condition is fulfilled without 
specifying  $T(t,r)$. Therefore, the function $T(t,r)$ in 
(\ref{gf}) remains arbitrary, which can probably be traced to a some kind 
of hidden gauge invariance. 

In case {II}  no restrictions on the coefficients $b_k$ arise,
so that this case is generic.  The coefficients $a,b,c$ can be obtained by 
comparing the line element $ds_f^2$ in (\ref{ansatz}) with that in (\ref{gf}), 
which gives 
\be
a^2-c^2=\frac{{\cal D}\dot{T}^2}{\Sigma} ,~~~~
b^2-c^2=\Sigma\left(\frac{u^2}{\cal D} -{\cal D}T^{\prime 2} \right),~~~~
c(a+b)={\cal D}\dot{T}T^\prime\,.
\ee
Resolving these relations with respect to $a,b,c$ and inserting the result to (\ref{cons2})
yields the equation,  
\be                           \label{cond-fin}
\frac{\cal D}{\Sigma}\,\dot{\cal T}^2
+\frac{\Sigma\cal D}{\Sigma-\cal D}\,{\cal T}^{\prime 2}=1\,,
\ee
with $T=u{\cal T}$.
A simple solution can be 
obtained by separating the variables, 
\be
{\cal T}=t+\int \frac{dr}{\Sigma}-\int\frac{dr}{\cal D}\,\equiv
t+r^\ast_\Sigma-r^\ast_{\cal D}.
\ee
One can think that 
this solution is singular, since the tortoise coordinate 
$r^\ast_{\Sigma}$ diverges at the black hole and cosmological horizons,
 where $\Sigma$
vanishes.  
However, introducing the light-like coordinate
\be
V=t+r^\ast_\Sigma={\cal T}+r^\ast_{\cal D}\,,
\ee
both metrics can be written in the Eiddington-Finkelstein form 
\bea 
ds_g^2&=&-\Sigma dV^2+2dVdr+r^2d\Omega^2\,,  \nonumber \\
\frac{1}{u^2}\,ds_f^2&=&-{\cal D} dV^2+2dVdr+r^2d\Omega^2\,,
\eea 
from where it is obvious that the solution is regular.  
This solution is valid for 
all values of the parameters $b_k$. All the above  
solutions have been obtained in the ghost-free bigravity context 
in \cite{Volkov:2012wp} (see also \cite{Comelli:2011wq}), 
but in fact solutions of this type were considered already 
long ago 
in the generic f-g bigravity theories 
\cite{Salam:1976as,Isham:1977rj,Berezhiani:2008nr}. 
The generalization for a nonzero electric charge was considered in 
\cite{Babichev:2014fka}. 

Since the f-metric becomes flat for $\eta\to 0$,
the solutions describe in this limit black holes in the dRGT massive gravity. 
In this context
they were studied in Refs.\cite{Nieuwenhuizen:2011sq,Berezhiani:2011mt} 
for the special case I, and in Refs.\cite{Koyama:2011xz,Koyama:2011yg}
for the generic case II. 
These solutions and their generalization for a nonzero electric charge 
\cite{Nieuwenhuizen:2011sq,Berezhiani:2011mt,Cai:2012db} exhaust all static,
spherically symmetric black holes in the dRGT theory.

\section{Hairy black holes, lumps, and stars}

Black holes considered in the previous two sections are described by the 
known GR metrics. New black holes 
are obtained in the case where  
the two metrics are simultaneously diagonal \cite{Volkov:2012wp}, 
\bea
ds_g^2&=&N^2dt^2-\frac{dr^2}{\Delta^2}-r^2d\Omega^2,~~~~
ds_f^2=A^2 dt^2-\frac{U^{\prime 2}}{Y^2} dr^2-U^2d\Omega^2\,.
\eea
Here 
$N,\Delta,Y ,U,A$ are 5 functions of $r$ which fulfill  
the equations 
\bea                              \label{eqs}
G^0_0&=&m^2{\cos^2\eta}\, T^0_0,~~~~~~~~
G^r_r=m^2\,{\cos^2\eta}\, T^r_r,~~\nonumber \\
{\cal G}^0_0&=&m^2\,{\sin^2\eta}\, {\cal T}^0_0,~~~~~~~
{\cal G}^r_r=m^2\,{\sin^2\eta}\, {\cal T}^r_r,\nonumber \\
{T^r_r}^\prime &+&\frac{N^\prime}{N}\,(T^r_r-T^0_0)
+\frac{2}{r}(T^\vartheta_\vartheta-T^r_r)=0.   
\eea 
The simplest solutions are obtained if
$f_{\mu\nu}=C^2g_{\mu\nu}$, where  $g_{\mu\nu}$ fulfills (\ref{g+f})
while  $C,\Lambda_g(C)$ are defined by 
(\ref{Lmbd}),(\ref{LAMC}). Since $\Lambda_g$ can 
be positive, negative, or zero, there are the
Schwarzschild, Schwarzschild-de Sitter, and 
Schwarzschild-AdS  black holes.  
Let us call them background black holes. 

More general solutions are obtained by numerically integrating Eqs.(\ref{eqs}). 
It turns out \cite{Volkov:2012wp} that the equations for the three amplitudes 
$\Delta,Y,U$ comprise a closed system. Its
 local solution near the horizon,  
\be                      \label{local}
\Delta^2=\sum_{n\geq 1}a_n(r-r_h)^n,~~
Y^2=\sum_{n\geq 1}b_n(r-r_h)^n,~~
U={ u}r_h+\sum_{n\geq 1}c_n(r-r_h)^n, \nonumber
\ee
contains {only one free parameter ${ u}$}=$U(r_h)/r_h$, which 
is the ratio of the horizon radius 
measured by $f_{\mu\nu}$ to that measured by $g_{\mu\nu}$. 
The horizon is common for both metrics,
in addition, its surface gravities and temperatures 
determined with respect to both metrics are the same \cite{Deffayet:2011rh}. 

Choosing a value of $u$ and integrating numerically the 
equations from $r=r_h$ towards large $r$, the result is as follows \cite{Volkov:2012wp}. 
If $u=C$ where $C$ is a root of the algebraic equation (\ref{LAMC}),
 then the solution 
is one of the background black holes. If $u=C+\delta u$ 
then one can expect the solution to be the   
background black hole slightly deformed by a massive graviton `hair' localized
in the horizon vicinity. 
\begin{figure}[th]
\hbox to \linewidth{ \hss

	\resizebox{8cm}{5cm}{\includegraphics{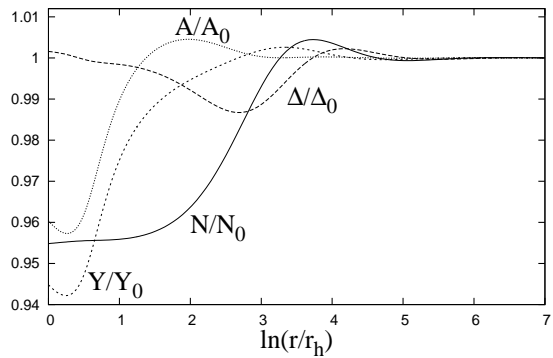}}

\hspace{1mm}
\hss}
\caption{{\protect\small 
Hairy deformations of the Schwarzschild-AdS background, where $A_0,N_0,\Delta_0,Y_0$
correspond to the undeformed solution. 
    }}
\label{Fig5}
\end{figure}
This is indeed confirmed for the Schwarzschild-AdS type 
solutions ($\Lambda_g<0$),
which can support a short massive hair and show 
deviations from the pure Schwarzschild-AdS
in the horizon vicinity, but far away from the   
horizon  the deviations tend to zero  (see Fig.\ref{Fig5}).
\begin{figure}[th]
\hbox to \linewidth{ \hss

	\psfrag{y}{$U^\prime$}
	\resizebox{8cm}{5cm}{\includegraphics{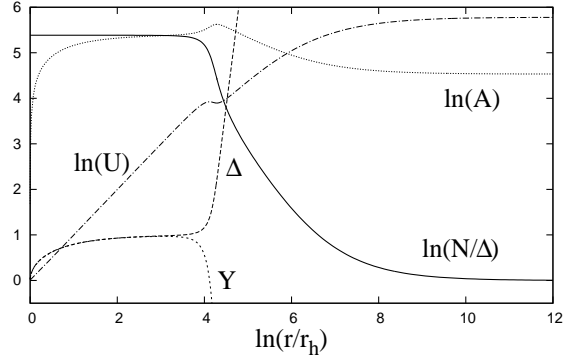}}
	
\hspace{1mm}
\hss}
\caption{{\protect\small 
Hairy deformations of the Schwarzschild background. 
    }}
\label{Fig5a}
\end{figure}
Therefore, there are asymptotically AdS hairy black holes in the theory. 

However, the procedure goes differently for $\Lambda_g\geq 0$. 
When one deforms the Schwarzschild background by setting $u=r_h+\delta u$, 
then the solutions first stay very close to 
Schwarzschild. However, at large $r$ they deviate away  
and show a completely different asymptotic behavior at infinity,
characterized by a quasi-AdS g-metric and a compact f-metric \cite{Volkov:2012wp}. 
Therefore, the only asymptotically flat black hole one finds 
is the pure Schwarzschild, while its hairy deformations loose the asymptotic flatness.  
Similarly, trying to deform the Schwarzschild-de Sitter background produces 
a curvature singularity at a finite
proper distance away from the black hole horizon,
hence the only asymptotically de Sitter black hole is the pure 
Schwarzschild-de Sitter. 

The conclusion is that there are hairy black holes in the theory, but they 
are not asymptotically flat. The following argument helps to understand this. 
Let us {\sl require} the 
solution to be asymptotically flat. Then one should have at large $r$ 
\bea                                      \label{infty}
\Delta&=&1-\frac{A\sin^2\eta}{r}+B\cos^2\eta \,\frac{mr+1 }{r}\,e^{-mr}+\ldots,\nonumber \\
U&=&r+B\,\frac{m^2r^2+mr+1 }{m^2r^2}\,e^{-mr}+\ldots, \nonumber \\
Y&=&1-\frac{A\sin^2\eta}{r}-B\,\sin^2\eta\,\frac{1+mr}{r}\,e^{-mr}+\ldots \,,
\eea
where $A,B$ are integration constants. 
Suppose that one wants to find black hole solutions with this asymptotic 
behavior using the multiple shooting method. In this method 
one tries to match the asymptotics (\ref{local}) and (\ref{infty}) 
by integrating the equations  starting from the horizon towards large $r$,
and at the same time starting from infinity towards small $r$.  
The two solutions should match at some    
intermediate point,  which gives three 
matching conditions   for
$\Delta,Y,U$. These conditions should be  
fulfilled by adjusting the free parameters
$A,B,u$ in Eqs.(\ref{local}),(\ref{infty}).  
Solutions of this problem may exist 
at most for discrete sets of values of $A,B,u$,
hence one cannot vary continuously  the horizon parameter $u$. 
Therefore, there 
could be no continuous, asymptotically flat  hairy deformations of
the Schwarzschild solution. However, this does not exclude isolated solutions,
and in fact they exist,  
but to find them requires a good initial guess for $A,B,u$.

It is interesting to see what happens to the hairy black holes 
when one changes the horizon radius $r_h$. 
It turns out that in the $r_h\to 0$ limit, where the black hole disappears, 
its `hair' survives and becomes 
a static `lump' made of massive field modes. Such lumps
are described by globally
regular solutions for which the event horizon is replaced by a regular center 
at $r=0$, while at infinity the asymptotic behavior is the same as for the 
black holes \cite{Volkov:2012wp}.  None of the lumps are asymptotically flat. 
Neither lumps no hairy black holes admit the dRGT
limit, they exist only in the bigravity theory. 

\begin{figure}[th]
\hbox to \linewidth{ \hss

	\resizebox{8cm}{5cm}{\includegraphics{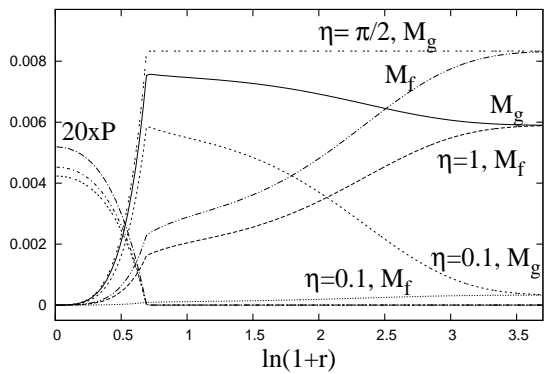}}

\hspace{1mm}
\hss}
\caption{{\protect\small Profiles of the  
asymptotically flat star solution sourced by a regular   
matter distribution. 
    }}
\label{Fig6}
\end{figure}

It is worth mentioning in this context that there are
asymptotically flat solutions with  a matter \cite{Volkov:2012wp}. 
Such solutions describe regular stars, and for them
one can take limits where one of the two metrics becomes flat.    
Suppose that
the f-sector is empty, while the g-sector contains 
$T^{[{\rm m}]\mu}_{~~~~~\nu}={\rm diag}[-\rho(r),P(r),P(r),P(r)]$ with 
$\rho=\rho_\star\theta(r_\star-r)$, corresponding to 
a `star' with a constant density $\rho_\star$ and a radius $r_\star$.  
Adding this source to the field equations (\ref{eqs}) and 
assuming a regular center  at $r=0$,
one finds solutions for which both metrics approach Minkowski metric at infinity
according to (\ref{infty}). 
Introducing the mass functions $M_g,M_f$ via 
$g^{rr}=\Delta^2=1-2M_g(r)/r$ and  $f^{rr}=Y^2/U^{\prime 2}=1-2M_f(r)/r$, 
one finds that $M_g,M_f$ rapidly increase inside the star, while outside
they approach the same asymptotic value $M_g(\infty)=M_f(\infty)\sim\sin^2\eta$
(see Fig.\ref{Fig6}). 
For $\eta=\pi/2$ the g-metric is coupled only to the matter and is described
by the GR Schwarzschild solution, $M_g(r)=\rho_\star r^3/6$ for $r<r_\star$ and 
$M_g(r)=\rho_\star r_\star^3/6\equiv M_{\rm ADM}$ for $r>r_\star$. 
For $\eta<\pi/2$ the star mass $M_{\rm ADM}$ is partially screened by the negative 
graviton energy. For $\eta=0$ (dRGT theory) the f-metric becomes flat, so that $M_f=0$,
while $M_g$ asymptotically approaches zero and the star mass
is totally screened, because the massless graviton decouples and there could be
no $1/r$ terms in the metric.  

If the graviton mass is very small, then the $m^2T^\mu_{~\nu}$ contribution 
to the equations is small as compared to $T^{[{\rm m}]\mu}_{~~~\nu}$, and
for this reason   
$M_g$ rests approximately constant for
$r_\star<r<r_{\rm V}\sim (M_{\rm ADM}/m^2)^{1/3}$. This illustrates the 
Vainshtein mechanism of recovery of General Relativity in a finite region
\cite{Vainshtein:1972sx}. This mechanism has also been confirmed 
by the numerical analysis 
within the generic massive gravity theory with the BD ghost 
\cite{Babichev:2009jt,Babichev:2010jd},
and also in the dRGT theory \cite{Gruzinov:2011mm}.
The approximate analytical solutions in the weak field limit 
were considered in Refs.\cite{Koyama:2011xz,Koyama:2011yg,Sbisa:2012zk} 
within the dRGT theory
and in Ref.\cite{Babichev:2013pfa} within the bigravity theory.

\section{Black hole stability and new hairy black holes}
As discussed in Section \ref{GR}, if the two metrics coincide, 
$
g_{\mu\nu}=f_{\mu\nu}
$, then 
the bigravity theory reduces to the vacuum GR, hence 
one can choose the Schwarzschild metric as a solution.
This  solution is known to be linearly stable in the GR context,
but one can wonder if it is stable also within the bigravity theory. 
Let us consider small perturbations around this solution, 
\be
g_{\mu\nu}= g^{\rm (0)}_{\mu\nu}+\delta g_{\mu\nu},~~~~
f_{\mu\nu}= g^{\rm (0)}_{\mu\nu}+\delta f_{\mu\nu}\,,
\ee
where $g^{\rm (0)}_{\mu\nu}$ is the Schwarzschild metric.
If one sets $\delta g_{\mu\nu}=\delta f_{\mu\nu}$, then the GR result will be 
recovered. However,  the perturbations of the 
two metrics need not be the same in general. 
Linearizing the bigravity field equations with respect 
to the perturbations, it turns out that the linear combinations 
\be
h_{\mu\nu}=\cos\eta\,\delta g_{\mu\nu}+\sin\eta\,\delta f_{\mu\nu},~~~
h^{\rm (0)}_{\mu\nu}=\cos\eta\,\delta f_{\mu\nu}-\sin\eta\,\delta g_{\mu\nu}
\ee 
decouple from each other and can be identified with the massive and massless
gravitons, respectively. Equations for the massless graviton are the same as in GR,
while for the massive graviton one obtains \cite{Babichev:2013una}
\bea                                  \label{GL}
&&\stackrel{(0)}{\Box} h_{\mu\nu}+2\stackrel{(0)}{R}_{\mu\alpha\nu\beta}h^{\alpha\beta}=
{m^2} h_{\mu\nu},~~~~~~~~~~~~~~~     \\
&&\stackrel{(0)}{\nabla}_\mu h^\mu_\nu=0,~~~h^\mu_\mu=0. \nonumber                
\eea
An interesting observation 
\cite{Babichev:2013una}
is that these equations have exactly the same structure as those 
describing perturbations of the black strings -- Schwarzschild black holes
uplifted to five spacetime dimensions.  At the same time, it is known that 
the black strings are prone to the Gregory-Laflamme instability
\cite{Gregory:1993vy}. Specifically, setting 
$
h_{\mu\nu}=e^{i{\omega} t}H_{\mu\nu}(r,\vartheta,\varphi), 
$
it turns out that Eqs.(\ref{GL}) admit a bound state solution with 
$
{\omega^2}<0 
$
in the  spherically-symmetric sector, 
provided that \cite{Brito:2013wya}
\be                                          \label{bound}
mr_h=\frac{\mbox{black hole radius}}{\mbox{graviton's Compton length}}<0.86.
\ee
It follows that small black holes are unstable,
since the frequency $\omega$ is imaginary and so the perturbations grow in time 
\cite{Babichev:2013una}. The condition of smallness is not crucial, since 
all usual black holes are small compared to the Hubble radius and so 
fulfill the bound (\ref{bound}), so that all of them 
should be unstable. On the other hand, since the frequency $|\omega|\propto m$, 
this instability is very mild, as it needs a Hubble time $\sim 1/m$ to develop. 
Therefore, even if real astrophysical black hole were described by the bigravity
theory, their instability would be largely irrelevant and they would actually 
be stable for all practical purposes over a cosmologically long period of time.  

A similar instability was found also for the Kerr black holes 
\cite{Brito:2013wya} and for the Schwarzschild-de Sitter black holes 
with proportional metrics described in Section \ref{GR} \cite{Brito:2013yxa}. 
Interestingly, it was found  in the latter case that the instability disappears
in the partially massless limit, where the graviton mass is related to the 
cosmological constant as $m^2=2\Lambda/3$ \cite{Brito:2013yxa}.  

As discussed in Section \ref{off} above, 
the Schwarzschild-de Sitter solution  
in the bigravity theory can exist also in a different version, 
for which the two metrics are not simultaneously diagonal. 
The linear stability of this solution was studied 
with respect to all possible perturbations, but only in the restricted case (\ref{cons1})
\cite{Kodama:2013rea}, and also in the generic case  (\ref{cons2}),
but only with respect to spherically symmetric perturbations \cite{Babichev:2014oua}. 
In both cases the solution was found to be stable. 

Getting back to the unstable Schwarzschild black holes, it turns out that 
their instability can be used to find new black holes which support hair and 
which are asymptotically flat. As was explained above, 
asymptotically flat solutions subject to the boundary conditions 
(\ref{local}),(\ref{infty}) may exist, but to find them requires to   
fine-tune the parameters $A,B,u$ in (\ref{local}),(\ref{infty}), for which
an additional information is needed. Now, the existence of the black hole 
instability provides such an information \cite{Brito:2013xaa}. 

Indeed, Eqs.(\ref{GL}) admit solutions with $\omega^2<0$ only for 
$mr_h<0.86$, while for $mr_h>0.86$ all solutions have $\omega^2>0$. 
This means that for $mr_h\approx 0.86$ there is a zero mode: a static 
solution of (\ref{GL}) with $\omega=0$.  This zero mode can be viewed as 
approximating a new black hole solution which exists for  $mr_h<0.86$ and which 
merges with the Schwarzschild solution for $mr_h\approx 0.86$. 
Close to the merging point the deviations of the 
new solution from the Schwarzschild are small and can be described by the linear theory. 
Therefore, one can use the linear zero mode  to read-off the values
of the parameters $A,B,u$ in (\ref{local}),(\ref{infty}), after which one can iteratively decrease 
$r_h$ to obtain the `fully-fledged' non-perturbative hairy black holes. 
This was done in Ref.\cite{Brito:2013xaa}.

The conclusion is that there are asymptotically flat black holes with a massive hair 
in the bigravity theory. However, it seems that their parameter $mr_h$ 
cannot be too small (unless for $c_3=-c_4=2$) \cite{Brito:2013xaa}, 
which means that these black holes are cosmologically large, their size being
comparable with the Hubble radius.   
Such solutions are unlikely to be relevant. 

All described above black holes have been obtained in the 
theory either without a matter source or in the theory with an electromagnetic field. 
At the same time, the perturbative analysis of Ref.\cite{Deser:2013rxa} 
predicts that hairy black holes should generically 
exist in the massive gravity theory coupled to a 
matter with a non-vanishing trace of the energy-momentum tensor.
It would be very interesting to test this prediction by fully non-perturbative calculations.

\section{Concluding remarks}
Summarizing the above discussion, 
all possible static, spherically symmetric black holes in the dRGT massive gravity theory are described 
by the Schwarzschild-de Sitter metrics. They belong to
the type studied in Section \ref{off} and they are
probably stable. 
One may wonder why one does not find asymptotically flat black holes. 
However, our universe is actually in the de Sitter phase, and 
the main motivation for considering theories with massive gravitons is to 
describe this fact. Hence, it is not astonishing that the solutions are not 
asymptotically flat.

One finds more solutions in the bigravity theory, as for example  
the hairy black holes.  However, these seem to be not very relevant,
since they are either asymptotically AdS, which contradicts the observations, 
or they are too large. There are also asymptotically flat or asymptotically 
de Sitter black hole solutions, but they are unstable. 
However, they can describe astrophysical black holes, 
since the instability takes cosmologically long times to develop.  
One can also wonder what these black holes decay to, and one possibility  is that 
their instability actually implies that there is
a slow accretion of massive graviton modes to the horizon \cite{Mirbabayi:2013sva}. 
If this is true, then the black holes should be almost 
exactly Kerr (Kerr-de Sitter), apart from small corrections in the near-horizon 
region where the accretion takes place.

Some aspects of the graviton mass can be captured within a simplified 
description in the context of the Galileon theory \cite{Nicolis:2008in}.
This is essentially the General Relativity coupled to a self-interacting 
scalar field that mimics the scalar polarization mode of the massive graviton.   
It turns out that black holes in these theory are described by the GR metrics
\cite{Babichev:2013cya}, and 
a no-hair theorem can be proven in this case \cite{Hui:2012qt}.

To recapitulate, even if the gravitons are indeed massive, this would be hard to 
detect by observing black holes.

\vspace{3 mm}

This work was partly supported by the Russian Government Program of Competitive Growth 
of the Kazan Federal University. 






\end{document}